# Machine Learning for Impurity Charge-State Transition Levels in Semiconductors from Elemental Properties Using Multi-Fidelity Datasets

Maciej P. Polak,[1, a)] Ryan Jacobs,[1] Arun Mannodi-Kanakkithodi,[2] Maria K. Y. Chan,[3, b)] and Dane Morgan[1, c)]
[1)] *Department of Materials Science and Engineering, University of Wisconsin-Madison, Madison, WI 53706-1595, USA*
[2)] *School of Materials Engineering, Purdue University, West Lafayette, IN, USA*
[3)] *Center for Nanoscale Materials, Argonne National Laboratory, Lemont, IL 60439, USA*

Quantifying charge-state transition energy levels of impurities in semiconductors is critical to understanding and engineering their optoelectronic properties for applications ranging from solar photovoltaics to infrared lasers. While these transition levels can be measured and calculated accurately, such efforts are time consuming and more rapid prediction methods would be beneficial. Here, we significantly reduce the time typically required to predict impurity transition levels using multi-fidelity datasets and a machine learning approach employing features based on elemental properties and impurity positions. We use transition levels obtained from low-fidelity (i.e. LDA or GGA) density functional theory (DFT) calculations, corrected using a recently-proposed modified band alignment scheme which well-approximates transition levels from high-fidelity DFT (i.e. hybrid HSE06). The model fit to the large multi-fidelity database shows improved accuracy compared to models trained on the more limited high-fidelity values. Crucially, in our approach when using the multi-fidelity data, high-fidelity values are not required for model training, significantly reducing the computational cost required for training the model. Our machine learning model of transition levels has a root mean squared (mean absolute) error of 0.36 (0.27) eV versus high-fidelity hybrid functional values, when averaged over 14 semiconductor systems from the II-VI and III-V families. As a guide for use on other systems, we assessed the model on simulated data to show the expected accuracy level as a function of band gap for new materials of interest. Finally, we use the model to predict a complete space of impurity charge-state transition levels in all zincblende III-V and II-VI systems.

## I. INTRODUCTION

Impurities and their charge-state transition levels i.e. the Fermi level at which the stable charge state of a defect changes (illustrated in Fig. 1), play a crucial role in determining the properties of a semiconducting material. Transition levels within the band gap influence the optical and electronic properties of the system. Defect transition levels may serve as traps for carriers and, thus, changing the conductivity, or act as non-radiative and radiative recombination centers, affecting device efficiency. Although experimental techniques such as deep level transient spectroscopy (DLTS)[1] are able to measure the presence of transition levels, they lack the ability to identify the underlying impurity or defect type, and they require special sample preparation. Although micro- scopic techniques such as cross-sectional scanning tunneling microscopy (STM)[2] or scanning transmission electron microscopy (STEM)[3] are able to provide additional insight into defect characterization, currently, there is no robust experimental technique which is able to determine the charge-state transition levels of a specific impurity in a semiconductor.

First-principles computational methods such as Density Functional Theory (DFT) can provide predictions for a number of defect properties, including charge-state transition levels[4]. Such theoretically obtained results not only provide valuable insight on their own, but can be, and often are, used in conjunction with experimental techniques to help identify the observed defects or impurities and provide more in-depth description of the physical mechanisms involved in defect engineering, such as materials processing schemes to enhance or mitigate the presence of certain defect types.

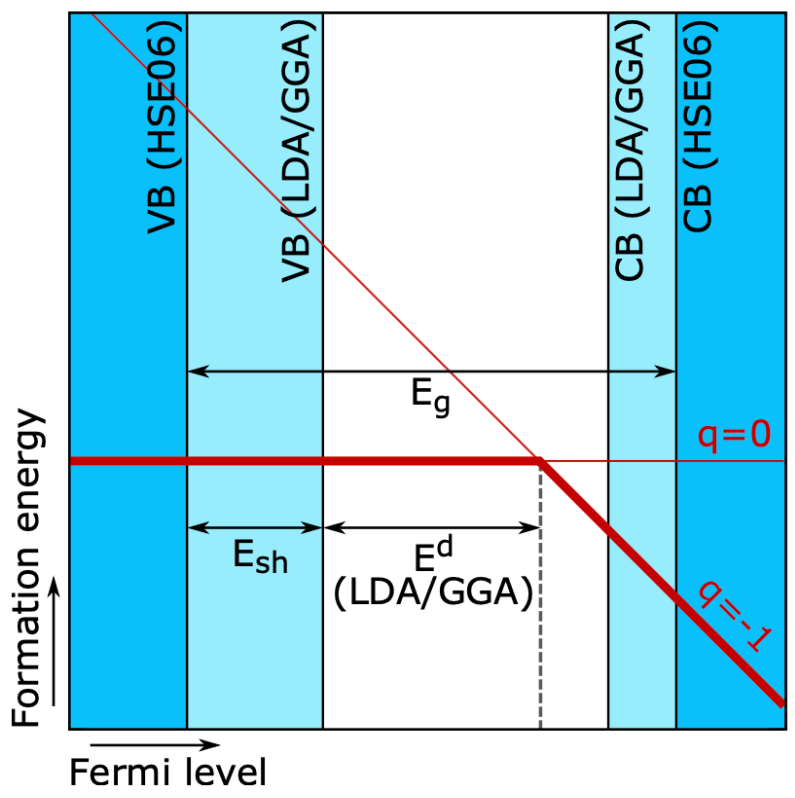

FIG. 1. A schematic picture illustrating charge state transition levels (defect levels) as intersections of lines representing formation energies for different charge states $q$ as a function of Fermi energy. CB and VB stand for conduction and valence band edges, respectively. Parameters necessary for modified band alignment (MBA) are also indicated: $E_{sh}$ - shift of the valence band between semilocal and hybrid functional, $E^d$ - semilocal defect level, and $E_g$ - the band gap.

However, for such calculations to be sufficiently accurate, the use of large supercells and high-fidelity computational approaches such as GW[5] or hybrid functionals (e.g. Heyd-Scuseria-Ernzerhof (HSE06)[6]) are necessary, making studies very computationally expensive and thus often restricted to a small number of material compositions or defect types. While

[a)]Electronic mail: mppolak@wisc.edu
[b)]Electronic mail: mchan@anl.gov
[c)]Electronic mail: ddmorgan@wisc.edu

high-fidelity calculations are certainly possible for a targeted and specific study, developing large databases of high-fidelity values is generally impractical due to the large computational expense. This limitation makes it difficult to develop adequate training databases of high-fidelity values for machine learning (ML) models.

In recent years, the use of ML in materials science for the purpose of predicting materials properties has been heavily explored and highly successful, with models developed to predict properties such as bulk stability, superconducting critical temperature, formability of metallic glasses, electronic bandgap, migration barriers for impurity diffusion, thermoelectric figure of merit, and effective charge for electromigration, to name just a few[7–17]. In particular, the concept of multi-fidelity machine learning, where different levels of approximations (fidelity) are combined in then training data and model design, have been proposed. This approach may allow for the use low-fidelity (therefore, cheaper to obtain) values to predict results at a high-fidelity level, potentially reduc- ing the computational effort needed for training of the ML model. This may be of particular value as more complete high-throughput low-fidelity databases become available[18]. Multi-fidelity methods have been shown to be effective for predicting materials properties such as band gaps[19,20], dopant formation energies[21], crystal structures, and bulk modulus, as well as for parameterizing interatomic potentials[22–24].

In the context of predicting impurity transition levels in semiconductors, developing an ML model is very desirable as it may provide rapid predictions without having to develop a large database of computationally expensive DFT calculations. In this work, we use multi-fidelity datasets to demonstrate an alternative ML approach that is capable of nearly instantaneously predicting impurity charge-state transition levels using only elemental and one-hot encoded features and without the need to perform additional DFT calculations. We show that an increased accuracy can be achieved when utilizing a larger and lower quality dataset for training instead of a smaller and more accurate one. It is worth to men- tion that the meaning of the term multi-fidelity in our ap- proach is slightly different than typically used. Here we im- prove the fidelity of our low-fidelity GGA and LDA training dataset by applying a correction that brings the low-fidelity values close to the high-fidelity HSE06 ones, and then ver- ify the accuracy of the trained model against true high-fidelity HSE06 results, therefore eliminating the necessity of includ- ing the true high-fidelity values in training. Although we use the terms low/high-fidelity to describe quantities calculated with semilocal (GGA/LDA) and hybrid functional (HSE06) respectively, the conclusions drawn from this research may be applicable more generally to multi-fidelity datasets, where different reasons than the functionals used account for varying levels of fidelity. That terminology is therefore used alongside the names of the functionals to convey the potential generality of the results. The ML models constructed in this work have been made possible by two recent studies. First, it was recently demonstrated from the work of Mannodi-Kanakkithodi et al.[25] that ML approaches can be used to estimate the impurity transition levels in II-VI Cd-based semiconductors. In that work, Mannodi-Kanakkithodi et al. used random forest models with descriptors consisting of properties of the involved chemical elements, the lattice sites containing impurities, and, optionally, results of fast unit-cell DFT calculations. ML models were constructed to predict transition levels obtained from both low-fidelity (PBE) and high fidelity (HSE06) DFT calculations. A root mean squared error (RMSE) of 0.34 eV was obtained for ML-predicted PBE transition levels vs. DFT-PBE calculated transition levels, while an RMSE of 0.61 eV was obtained for ML-predicted HSE06 transi- tion levels vs. DFT-HSE06 calculated transition levels. Sec- ond, from the perspective of enhancing the accuracy of defect transition levels obtained from DFT, recent work from Polak et al.[26] formulated the modified band alignment (MBA) approach as an enhancement to the previously-used band alignment approach[27,28]. The MBA for defect levels is expressed as:

$$E^d_{MBA} = E^d + E_{sh} + \beta E_g - E^d \quad . \tag{1}$$

where $E^d_{MBA}$ is the MBA calculated defect level, $E^d$ is the local/semilocal defect level, $E_g$ is the band gap, $E_{sh}$ is a band alignment shift and $\beta = -0.14$ for the systems studied here. Figure 1 illustrates the necessary quantities, and the whole approach is described in detail in Ref.[26]. This MBA approach was found to improve the accuracy of defect transition levels obtained with local or semilocal functionals (i.e. low-fidelity LDA or PBE calculations) vs. values obtained from high-fidelity HSE06 calculations. In the work of Polak et al., an RMSE of 0.24 eV was obtained for the MBA-calculated transition levels vs. high-fidelity HSE06 calculated results.

To obtain an approximation of the level of expected accuracy by combining similar ML methods as used in Mannodi-Kanakkithodi et al. with the MBA approach of Polak et al., we add the above RMSE values in quadrature, resulting in an RMSE of 0.42 eV, which is about 32% lower than the value of 0.61 eV obtained for the direct ML prediction of HSE06 transition levels from Mannodi-Kanakkithodi et al.[25] This reduction in error, if realized, may be attributable to the larger amount of MBA corrected PBE vs. HSE06 data available for training. This quick approximation motivates the construction of the ML models in the present work, as it is now apparent the use of ML combined with the MBA has the potential to provide predictions of high-fidelity charge-state transition levels with significantly lower error than obtained with previous ML models.

Building from the results of these previous studies and the above error approximation, there is thus an opportunity to construct improved ML models using only low-fidelity training data, where this training data is modified using the MBA approach to more closely resemble the accuracy level of high-fidelity data. Such a multi-fidelity model could not only significantly improve the accuracy in predicting high-fidelity transition levels, but would also remove the necessity of performing the computationally expensive high-fidelity HSE06 calculations for the construction of the training data set. The goal of this work is to build a model that enables fast and inexpensive estimation of defect and impurity transition levels with a level of accuracy rivaling high-fidelity DFT calculations using feature set containing no information from DFT calculations, only readily available elemental properties. Such a model would make qualitative or even semi-quantitative insight into semiconductor impurity transition level properties easily obtainable to researchers from both experimental and computational backgrounds.

## II. DATASETS AND METHODS

The dataset used for this study consists of 2910 low-fidelity (LDA or GGA) impurity charge-state transition levels in 14 zincblende semiconducting systems. These include Cd-based II-VI semiconductors and their alloys (i.e. CdS, CdSe, CdTe, $CdS_{0.5}Se_{0.5}$, and $CdSe_{0.5}Te_{0.5}$) and all binary stoichiometric III-V systems (i.e. AlP, AlAs, AlSb, GaP, GaAs, GaSb, InP, InAs, and InSb). This dataset is used for the training, in- ternal validation, and optimization of the models. From the 2910 calculated systems, 896 were additionally studied with the high-fidelity hybrid HSE06 functional in order to test the final accuracy and performance of our ML models against ac- curate data. The exact list of the 2910 low-fidelity systems and the 896 high-fidelity ones can be found in Sec. VII. All of the DFT calculations were performed with the `VASP` code[29,30] on 64-atom supercells where the spurious electrostatic interac- tions were corrected with the Freysoldt-Neugebauer-Van de Walle (FNV) correction scheme[31,32]. The datasets used in training of the models included spin polarization and full ionic optimization, with symmetry constraints switched off. However, we did not explicitly move atoms to break the symmetry of a given atomic arrangement. This choice can lead to the relaxation algorithm failing to find some local minima[33] but was made to allow efficiency in the high-throughput calculations that enable this study. More details of the DFT calculations can be found in Mannodi-Kanakkithodi et al.[25] and Polak et al.[26]

The LDA/GGA results from these studies were corrected with the MBA scheme, which is described in details in Ref.[26]. That approach eliminated the possible inconsistency coming from mixing LDA and PBE calculations in the same dataset. These MBA-corrected low-fidelity transition levels were used as the target quantity to predict with ML models.

### A. Model selection

All ML model fits were performed using the Materials Simulation Toolkit for Machine Learning (MAST-ML)[34]. MAST-ML is a python-based toolkit built to enable easy auto- mated exploration of many ML model types and statistical assessments of model performance, and heavily leverages the widely-used scikit-learn package[35]. From an initial exploration of numerous canonical model types (e.g. kernel ridge regression, random forests, Gaussian process regression, etc.), we found that gradient boosting regression (GBR)[36] resulted in the best overall model performance in terms of lowest RMSEs from 10-fold cross validation (CV). In the 10-fold CV used here, the dataset consisting of 2910 datapoints is split into 10 equal randomized batches (folds) of 291 points and each batch is then used as validation while using the remaining 9 batches for training. We note here that 10-fold CV was used in this work to enable straightforward comparison of model performance compared to the work of Mannodi-Kanakkithodi et al., who also evaluated ML models with 10-fold CV[25].

### B. Feature selection and hyperparameter optimization

To find the optimal number of features for our ML model, a feature learning curve was calculated where the test mean absolute error (MAE) and RMSE scores were plotted as a function of number of selected features (Fig. S2 (a)), where the features were selected using the GBR ensemble feature importance ranking. Hyperparameter optimization was performed using the Bayesian search optimization method from the scikit-optimize package, which is integrated as part of the MAST-ML workflow. The GBR model hyperparameters and feature selection were carried out using a nested CV approach to remove possible bias from using all of the training data to either select the best feature set or best set of model hyperparameters. Specifically, for each of the 10 training sets from the top level CV, we perform an additional nested 10-fold CV and minimize that score to select the optimal hyperparameters and features. We then use those optimal hyperparameters and features to refit the GBR model and predict the top level validation data and assess prediction errors. We found that the features with highest importance remained constant across all training sets, were physically meaningful, and included properties such as the impurity charge states, one-hot encoded impurity position, ionization energies, electronegativities, and electrical conductivity. When the selected features between different training sets were compared, features with low importance often changed or changed order, which did not seem to noticeably influence the overall results. It was determined that there was no significant model improvement when more than 15 features were used, so the 15 features with the highest importance were chosen to be used in the final model. The learning curve is presented in Fig. S2 (a) and the list of fi- nal features and their importances can be found in Sec. VII. Additionally, to assess the potential for improvement of the model by using more training data, a data learning curve was calculated where the test MAE and RMSE were plotted as a function of number of training points (Fig. S2 (b)). The data learning curve shows that including more data in the train- ing may have the potential to further increase the accuracy of the model, where an additional approximately 1500 data points (an increase of about 50 %) may reduce the MAE and RMSE by about 50 meV, provided the established learning curve trend persists to higher amounts of training data. Such improvement of the model with additional training data is feasible as only relatively inexpensive low-fidelity DFT calculations are needed to generate additional training data.

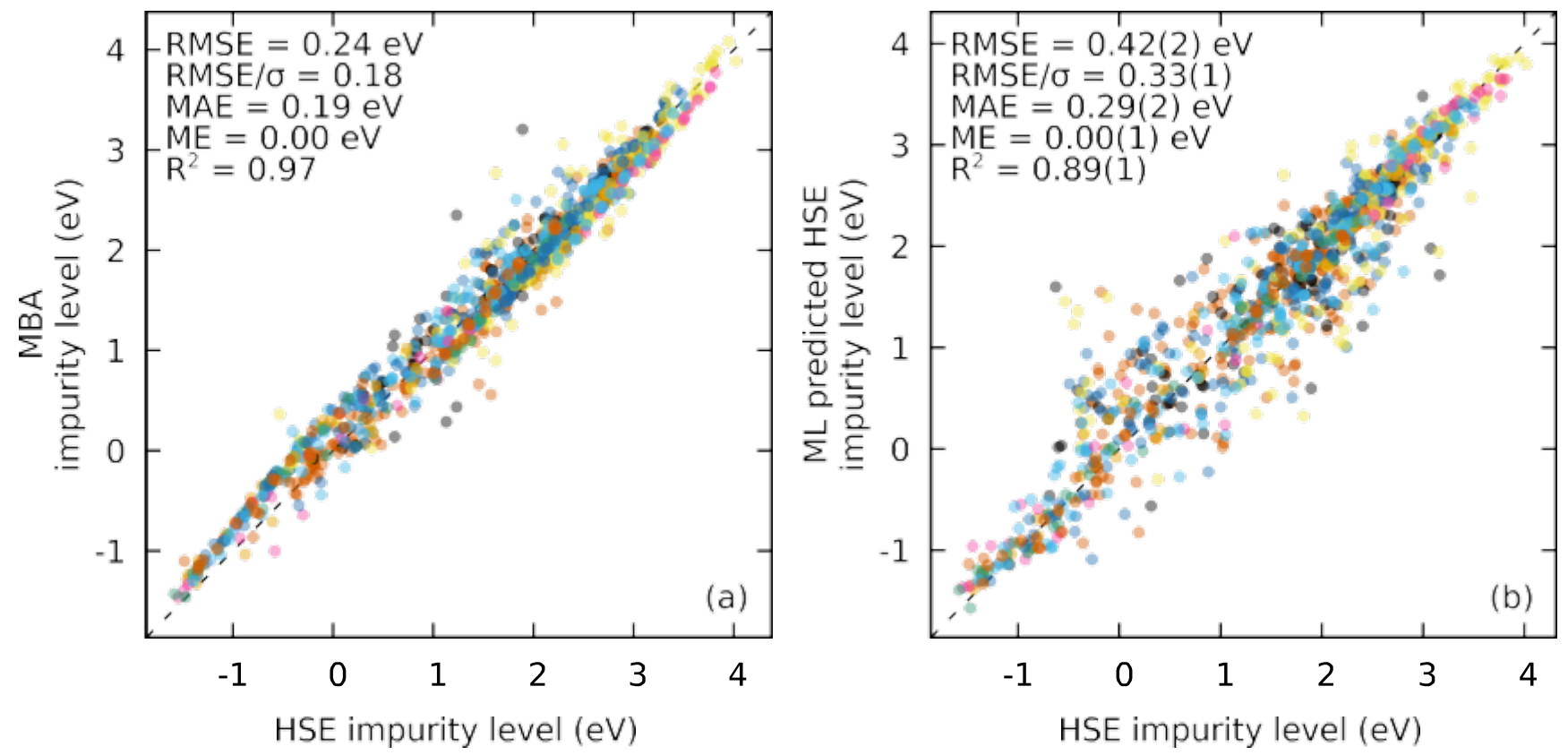

FIG. 2. Parity plots of 10-fold cross validation of different approaches to predicting charge-state transition levels. (a) MBA vs. HSE06, (b) ML predicted HSE06 vs. HSE06. Different point colors represent different systems (See Fig. S5).

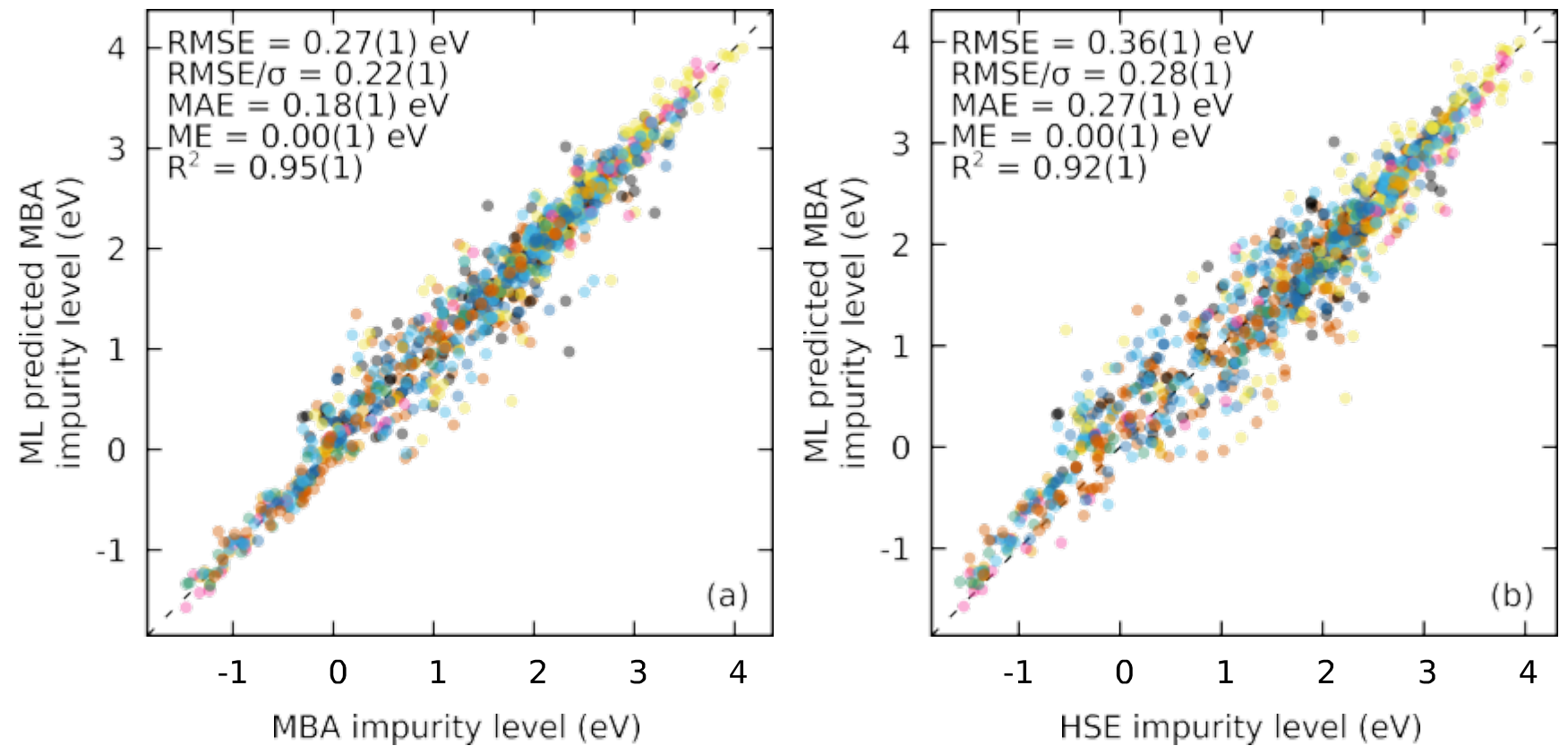

FIG. 3. Parity plots of 10-fold cross validation of the final model predicting charge-state transition levels. (a) ML predicted MBA vs. MBA, (b) ML predicted MBA vs. HSE06. Different point colors represent different systems (See Fig. S5.

## III. RESULTS AND DISCUSSION

In this study, we are working with an expanded dataset of II-VI and III-V semiconductor materials compared to the datasets examined in Mannodi-Kanakkithodi et al. and Polak et al. Thus, as a first step, we re-evaluate the predictive ability of MBA vs. HSE06 transition level values and ML-predicted HSE06 values, as shown by the parity plots in Fig. 2. In Fig. 2(a), we plot the MBA values of all II-VI and III-V materials for which corresponding HSE06 data is available in our database, which amounts to 896 materials. From Fig. 2(a), the MBA method results in an RMSE of 0.24 eV vs. HSE06 values, which is the same the previous RMSE value of 0.24 eV from Polak et al., which only used III-V compounds. Figure 2(b) contains the parity plot for ML-predicted HSE06 values, which has an RMSE of 0.42 eV. This value is lower than the RMSE of 0.61 eV from the work of Mannodi-Kanakkithodi et al., which we believe is primarily due to having a larger amount of HSE06 data in the present work. This scale of improvement is reasonable based on the HSE06 data learning curve presented in the work of Mannodi-Kanakkithodi et al., which showed an RMSE improvement of roughly 0.1 eV per 100 added data points. Given the present work uses roughly 200 more HSE06 data points, the improvement of 0.61 eV to 0.42 eV is thus expected considering this trend and the fact that the RMSE improvement per data point added will level off for larger dataset sizes.

Figure 3(a) contains a parity plot showing our ML model performance for predicting the MBA transition level values, where we have an RMSE of 0.27 eV. Note that the data shown in Figure 3(a) are the ML-predicted MBA values only for the materials for which HSE06 data is also available. An analogous plot showing the ML-predicted MBA for the full dataset is shown in Fig. S3 (a) in the Supporting Information, which showed a negligible difference in RMSE (0.26 eV (all data) and 0.27 eV (HSE06 only)). Re-evaluating the previously discussed approximate RMSE from combining the separate errors of MBA vs. HSE06 and ML-MBA vs. MBA, results in an RMSE of 0.36 eV (from adding errors in quadrature). This value is lower than the previous estimate of 0.42 eV and corresponding to an improvement of about 14% for the new multi-fidelity method, which in addition does not require access to the high-fidelity data for model training. Figure 3(b) presents a parity plot of the ML-predicted MBA vs. HSE06 transition state values, with an RMSE of 0.36 eV, the same as the expected RMSE of 0.36 eV. Supplemental Table SI contains the tabulated errors for the ML-predicted MBA vs. HSE06 values from Figure 3(b) for all data and across each of the 14 semiconducting systems examined here.

There are other methods, besides fitting the MBA values, that one may use to construct ML models to predict high fidelity transition levels. It was previously shown by Mannodi-Kanakkithodi et al. that including low-fidelity level (PBE) transition levels as a feature for predicting HSE06 results leads to a substantial improvement in the model accuracy[25]. While this method relies on using DFT calculations to first obtain the PBE transition levels, it suggests the possibility for application of the latent variable transfer learning technique[37], where in the first step, the low-fidelity (MBA or PBE) results are predicted from a machine learning model using elemental properties, and then the obtained results are used as an additional feature in predicting the high-fidelity values. The results of this approach are presented in Fig. S1 (b) and (c) of the Supporting Information. This transfer learning tech- nique does improve the prediction statistics compared to the direct prediction of HSE06 values (Fig. 2(b)). However, this approach does not provide a noticeable improvement over the multi-fidelity method proposed in this work. Furthermore, this approach requires HSE06 calculations, typically quite a large number if one wants to have a good training database, which is not true of our approach which uses only MBA modified LDA and GGA values. Therefore, the approach described in this work appears to be preferable.

The obtained error statistics of our multi-fidelity model are very encouraging considering the complexity of the physi- cal mechanisms behind impurity formation energies and their transition levels, and the fact that the model uses only basic elemental and one-hot encoded features. However, it can be easily observed, that with a RMSE (MAE) of 0.36 (0.27) eV, the model is not quantitatively useful for systems with low band gaps, where a charge-state transition level may easily be predicted to be a deep defect when it should be shallow, and vice versa. Note that we find errors are similar for lower and higher band gap materials, making the errors particularly problematic for systems with low band gaps. In addition, it is worth noting that for some defects and charge states, an LDA/GGA approach may incorrectly capture localization effects[27,38], which is not corrected for in any band alignment-like method, including MBA[26]. These cases are difficult to unambiguously identify in a high-throughput approach, and are therefore included in this model as well, influencing the statistics. Ref.[27] has demonstrated that large discrepancies are due to this issue, so it is most likely responsible for the outlier points in Fig. 2 (a), which, fortunately, are relatively infrequent; therefore, their influence on the final model is not significant. To help address this issue and provide more insight into the practical applicability of the present model, we have devised a sim- ple method for quantitative estimation of model applicability based on precision, recall and F1 scores[39] that enables the interested user to investigate the performance of the model as a function of the band gap of a hypothetical system of interest. Here, we have considered three different practical scenarios where this model can be used for assessment of transition level locations, based on the level of precision needed: (a) whether a transition level exists in the band gap, (b) whether a transition level resides in the upper or lower half of the band gap, and (c) whether a transition level resides in the upper, middle or lower third of the band gap. To illustrate the model util- ity as a function of band gap, we generated synthetic HSE06 transition levels for systems of band gaps varying from 0.25 eV to 6 eV. For simplicity, we assumed that the impurity transition levels are distributed uniformly within the band gap and up to $\pm 3\sigma$ (of the impurity levels) outside of the band gap. We then generated predicted transition levels with the residuals randomized with a Gaussian distribution but with the same error statistics (MAE and RMSE) of our ML-predicted MBA vs. HSE06 data from Fig. 3. This approach enables us to simulate the performance of the model for a wider range of materials of higher band gaps, where it is expected to perform better. The results of this estimation of model performance are shown in Fig. 4 as solid lines, where the precision, recall and F1 scores are presented as a function of the band gap, for the three different scenarios described above.

The assumption of the impurity transition levels being uniformly distributed within the band gap is simple and straightforward, but not necessarily true. To validate its applicability in this case, in a similar fashion as before we re-sampled the predictions of actual HSE06 impurity levels with the residuals randomized with a Gaussian distribution with the same statistics of our ML-predicted MBA vs. HSE06 data from Fig. 3. The results are plotted in Fig. 4 as points. The relatively good agreement of the points with the curves shows that the uniform distribution assumption does not alter the final results and is likely to be accurate for higher values of band gaps as well. Overall, Fig. 4 proves what was initially expected from the model, i.e. the performance is improved for systems with higher band gaps. While the results of the model will be of lower quantitative accuracy than a full suite of HSE06-level calculations, in many practical use cases, such quantitative accuracy may not be needed. For understanding the behavior of impurity transition levels in semiconducting systems, it is often most relevant whether the impurity transition level manifests within the band gap, as such a transition level may greatly influence the optical and electronic properties of the system. From Fig. 4(a), systems with band gaps of around 1 eV or higher have F1 scores above 75% for the scenario of whether a transition level resides within the band gap. Shrinking the region of testing to half, or a third of the band gap (panels (b) and (c)) reduces the performance scores. Therefore, whether the model is applicable to a particular problem has to be assessed based on the desired accuracy and material of interest.

### A. Model application

The whole space of binary III-V and II-VI zincblende semiconductors is composed of 18 systems (including those where zincblende is metastable but can be synthesized and excluding alloys). Considering 63 sensible impurity elements, 5 de-fect sites and 7 charge states (following the work of Mannodi-Kanakkithodi et al.[25]), the entire space is composed of 34,524 possible impurity charge-state transition levels.

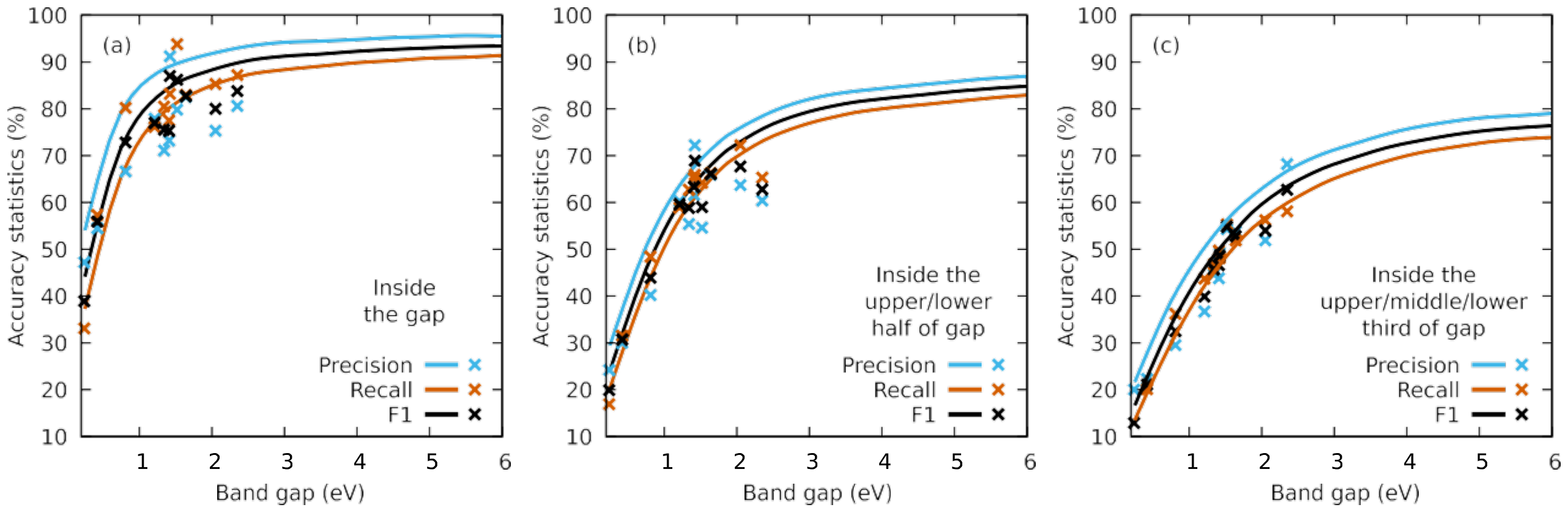

FIG. 4. Simulated precision, recall, and F1 score as a function of the band gap of the system, calculated for three different criteria of proper classification in: (a) the gap, (b) lower or upper half of the gap, (c) lower, middle or upper third of the band gap. Lines are calculated with the assumption that impurity levels are uniformly distributed within the gap, points use actual distribution of the impurity levels from the calculated data for all studied materials.

The 2910 transition levels included in the training and validation of the model covers less than 10% of this space. The model presented in this study should be capable of predicting the remaining 90%, the results of which are gathered in Sec. VII. Although the accuracy of the predictions vs. high-fidelity calculations or experiments is not currently known, the conclusions drawn in Sec. III from Fig. 4 should provide an adequate guide. Additional analysis of the results revealed that there were no outlying points or unreasonably low or high values. The histogram of the transition levels for the entire space is visualized in Fig. S4.

## IV. SUMMARY AND CONCLUSION

In this work, using multi-fidelity datasets, we have constructed a machine learning model capable of predicting charge-state transition levels of impurities in II-VI and III-V semiconducting systems with an RMSE (MAE) of 0.36 (0.27) eV compared to high-fidelity hybrid functional HSE06 DFT calculations. Although the accuracy may not be good enough for certain applications, where a more re- fined model may be needed, it allows for qualitative and semi-quantitative predictions, especially useful in high through- put screening for systems of potential interest. Our ap- proach demonstrates an alternative way to build accurate mod- els without the need to perform high-fidelity calculations for training. The fitted ML model allows for nearly instantaneous prediction of impurity levels without the need to perform additional DFT calculations. It uses the results of low-fidelity (PBE or LDA functional) calculations corrected with the modified band alignment method as a target and only elemental and one-hot encoded values as features. The present model trained on a large database of MBA corrected PBE and LDA calculations has higher accuracy in predicting validation HSE06 values than models that directly fit a smaller database of HSE06 values. It shows that using a much larger dataset of lower quality data for training may be superior to using a more accurate but smaller dataset, while at the same time be- ing computationally less expensive. This paves the way for further refinement of the model to obtain quantitatively accurate predictions of defect and impurity levels in semiconductors, and shows an alternative path to building ML models of materials properties using less expensive and more abundant data. To further assess the applicability of the method for practical applications, we provide guidance on classification of transition level positions within the band gap based on the value of the band gap, providing a useful assessment of model performance on new systems of interest. Finally, we use the model to directly predict the remaining 90% of the impurity charge-state transition levels not included in the training data, generating a large database of ML-predicted transition levels for all II-VI and III-V semiconductor systems.

## V. SUPPLEMENTARY MATERIAL

See supplementary material for supplementary figures S1-S5 and table SI.

## VI. ACKNOWLEDGMENT

We gratefully acknowledge support from the NSF Cyberinfrastructure for Sustained Scientific Innovation (CSSI) award No. 1931298. This work was performed, in part, at the Center for Nanoscale Materials, a U.S. Department of En- ergy Office of Science User Facility, and supported by the U.S. Department of Energy, Office of Science, under Contract No. DE-AC02-06CH11357. We acknowledge funding from the US Department of Energy SunShot program under contract # DOE DE-EE005956. This research used resources of the National Energy Research Scientific Computing Center, a DOE Office of Science User Facility supported by the Office of Science of the U.S. Department of Energy under Contract No. DE-AC02-05CH11231. A.M.K. acknowledges support from the School of Materials Engineering at Purdue University under account number F.10023800.05.002. We gratefully acknowledge the computing resources provided on Bebop, a high-performance computing cluster operated by the Laboratory Computing Resource Center at Argonne National Laboratory.

## VII. DATA AVAILABILITY STATEMENT

A spreadsheet of all the values used for model training, the values predicted with the use of the model, as well as a python pickle file with the saved scikit-learn model are openly available on figshare[40]. Raw data for the CdS systems is available at[41]. Since this model is most likely going to be used to predict new systems, in the downloadable model the entire dataset has been used in the feature selection and hyperparameters optimization, therefore, it is likely to have a slightly improved CV scores.